\documentclass[,final] {aipproc}
\usepackage{amssymb}
\usepackage{amsfonts}
\usepackage{amsmath}

\layoutstyle{8x11single}
\begin{document}

\title{Massive gravitational waves from the Cosmic Defect theory}


\author{N. Radicella}{
 address={Dipartimento di Fisica, Politecnico, Corso Duca degli Abruzzi 24, 10129 Torino and INFN, Sezione di Torino, Via Pietro Giuria 1, 10126 Torino}
}

\author{A. Tartaglia}{
 address={Dipartimento di Fisica, Politecnico, Corso Duca degli Abruzzi 24, 10129 Torino and INFN, Sezione di Torino, Via Pietro Giuria 1, 10126 Torino}
}

\begin{abstract}
The Cosmic Defect theory (CD), which is presented elsewhere in this conference, introduces in the standard Einstein-Hilbert Lagrangian an {\sl elastic} term accounting for the strain of space-time viewed as a four-dimensional physical continuum. In this framework the Ricci scalar acts as the kinetical term of the strain field whose potential is represented by the additional terms. Here we are presenting the linearised version of the theory in order to analyze its implications in the weak field limit. First we discuss the recovery of the Newtonian limit. We find that the typical static weak field limit imposes a constraint on the values of the two parameters (Lam\'e coefficients) of the theory. Once the constraint has been implemented, the typical gravitational potential turns out to be Yukawa-like. The value for the Yukawa parameter is consistent with the constraints coming from the experimental data at the Solar system and galactic scales. We then come to the propagating solutions of the linearised Einstein equations in vacuo, i.e. to gravitational waves. Here, analogously with other alternative or extended theories of gravity, the presence of the strain field produces {\sl massive} waves, where {\sl massive} (in this completely classical context) means subluminal.  Furthermore longitudinal polarization modes are allowed too, thus lending, in principle, a way for discriminating these waves from the plane GR ones.

\end{abstract}

\maketitle

\section{Framework}
The Lagrangian density we want to analyse is 
$$
\mathcal{L}=R+\frac{1}{2}C_{\mu\nu\rho\sigma}\varepsilon^{\mu\nu}\varepsilon^{\rho\sigma},
$$
where the second term on the r.h.s. is a potential term, in the form of an elastic potential whose field is the strain tensor $\varepsilon_{\mu\nu}\doteq \frac{1}{2}(g_{\mu\nu}-\eta_{\mu\nu})$, and the Ricci scalar is computed by means of the observed metric $g_{\mu\nu}$.\\ 
The  strain is ascribed to a cosmic point-like defect, this is why we called the theory the Cosmic Defect theory \cite{CD}
In our analogy we are interested in an isotropic medium; in this case the elastic constants take a simple 
form:
$$
C_{\mu\nu\rho\sigma}=\lambda \eta_{\mu\nu} \eta_{\rho\sigma} +\mu (\eta_{\mu\rho}\eta_{\nu\sigma}+
\eta_{\mu\sigma}\eta_{\nu\rho}).
$$
They now depends on two parameters only, the so-called Lam\'{e} coefficients $\lambda$ and $\mu$.\\
The Elastic potential then translates in 
$$
\mathcal{V}=\frac{1}{2}\left[\lambda \varepsilon^2+2\mu \varepsilon_{\mu\nu} \varepsilon^{\mu\nu}\right],
$$
where, following the structure of the elastic coefficients, the covariant version of the $\varepsilon$ tensor is obtained by lowering the indices with the total metric $g$.\\
The action integral is then
\begin{equation}\label{action}
S=\int d^4 x\left[R+\mathcal{V}\right]\sqrt{-g}.
\end{equation}
By varying the action in eq.(\ref{action}) w.r.t. the metric $g^{\mu\nu}$ that is the only dynamical field we obtain the Elastic Einstein Equations (EEE):
\begin{equation}\label{EEE}
G_{\mu\nu}=T^e_{\mu\nu}
\end{equation}
where 
$$
T^e_{\mu \nu }=\lambda \varepsilon \left[ \left( \frac{\varepsilon }{4}+\frac{
1}{2}\right) g_{\mu \nu }-\varepsilon _{\mu \nu }\right]\nonumber\\
 +\mu \left[
\varepsilon _{\mu \nu }+\frac{1}{2}g_{\mu \nu }\varepsilon _{\alpha \beta
}\varepsilon ^{\alpha \beta }-2\varepsilon _{\mu \alpha }\varepsilon _{\nu
}^{\alpha }\right].
$$

\section{Weak field limit}
In order to look for the gravitational waves of this theory we must linearize it around the Minkowski space-time, to which the theory reduces locally when the defect is not so strong. \\
It means that the metric can be written as
$$
g_{\mu\nu}=\eta_{\mu\nu}+\epsilon h_{\mu\nu}\quad \quad \epsilon\ll1 ,
$$
where the perturation metric $h_{\mu\nu}$ is driven by $\epsilon$.
The strain tensor becomes
$$
\varepsilon_{\mu\nu}=\frac{1}{2}\left[g_{\mu\nu}-\eta_{\mu\nu}\right]=\frac{1}{2}\epsilon h_{\mu\nu},
$$
where  we use the coordinate invariance of the theory in order to fix the flat metric to be the Minkowski one (i.e. $Diag(-1,1,1,1)$).
The strain tensor represents the perturbation itself.
Looking at the EEE, they are:
\begin{equation}\label{einstein}
G_{\mu \nu }=T_{e\mu \nu },
\end{equation}
where 
\begin{equation}
T_{e\mu \nu }=\lambda \varepsilon \left[ \left( \frac{\varepsilon }{4}+\frac{
1}{2}\right) g_{\mu \nu }-\varepsilon _{\mu \nu }\right]
 +\mu \left[
\varepsilon _{\mu \nu }+\frac{1}{2}g_{\mu \nu }\varepsilon _{\alpha \beta
}\varepsilon ^{\alpha \beta }-2\varepsilon _{\mu \alpha }\varepsilon _{\nu
}^{\alpha }\right].  
\label{tensore}
\end{equation}
First of all we should look at the linearized Einstein tensor, that, from
$$
R ^{\alpha}_{\beta\gamma\delta}\simeq\partial_{\gamma} \Gamma^{\alpha}_{\beta\delta}-\partial_\delta\Gamma^{\alpha}_{\beta\gamma},
$$
reduces to 
$$
G_{\alpha\beta}\simeq\frac{1}{2}\left[\partial_\gamma\partial_\beta h^{\gamma}_{\alpha}+\partial^\gamma\partial_\alpha h_{\beta\gamma}-\Box h_{\alpha\beta}-\partial_\alpha\partial_\beta h-\eta_{\alpha\beta}\partial_{\gamma}\partial^\delta h^{\gamma}_\delta +\eta_{\alpha\beta}\Box h\right],
$$
where indices are raised and lowered by means of the Minkowski metric and $h$ is the trace of the perturbation\footnote{It is worthwhile here to be precise about the sign convention on Riemann and Ricci tensors. 
In this paper we use the Riemann tensor written above, where the upper index is the first one, and we contract the first and the third index in Riemann to obtain Ricci.}. \\
The linearised Elastic energy-momentum tensor becomes
$$
T^e_{\mu\nu}\simeq\frac{\lambda}{2}\varepsilon \eta_{\mu\nu}+\mu \varepsilon_{\mu\nu}.
$$
The linearised EEE  in vacuo reduce to 
\begin{equation}\label{linEEE}
\begin{split}
\left[\partial_\gamma\partial_\beta h^{\gamma}_{\alpha}+\partial^\gamma \partial_\alpha h_{\beta\gamma}-\Box h_{\alpha\beta}-\partial_\alpha \partial_\beta h\right.&\left.-\eta_{\alpha\beta}\partial_{\gamma}\partial^\delta h^{\gamma}_\delta +\eta_{\alpha\beta}\Box h\right]\\
&-\mu\left(h_{\mu\nu}+\frac{\lambda}{2\mu}h \eta_{\mu\nu}\right)=0
\end{split}
\end{equation}
In order to investigate more deeply these equations we rewrite them by using their divergence and trace. The divergence, thanks to the contracted Bianchi identities, gives a relation between the divergence of the perturbation and that of its trace:
\begin{equation}\label{divergence}
h^{\mu\nu}_{;\nu}=-\frac{\lambda}{2\mu} h^{;\mu}.
\end{equation}
By tracing eq.(\ref {linEEE}) one gets 
\begin{equation}\label{trace}
2\left(1+\frac{\lambda}{2\mu}\right)\Box h-\mu\left(1+\frac{2\lambda}{\mu}\right) h=0.
\end{equation}
The trace of the perturbation becomes a dynamical field. This degree of freedom is always ghostlike, regardless the combination of the parameters \cite{boulware, deser90} (and it could be tachyonic, too): this makes the theory pathological from the quantum point of view. To avoid this behaviour we must constraint our parameters so that $1+\lambda/2\mu=0$. This choice reduces our linearised Elastic energy-momentum tensor to the Pauli-Fierz mass term \cite{fierz,pauli}, which actually leads to different predictions from those of General Relativity, no matter how small the graviton mass is \cite{carrera,deser}. This is what is called van Dam-Veltman-Zakharov discontinuity \cite{dam,zakharov}, whose existence, accordingly to some authors, can be ascribed to the explicit breaking of the gauge invariance by the mass term so that it can be cured if the mass is generated by the compactification of a higher dimensional theory \cite{deffayet02} or to the evaluation of the linearised theory around flat backgrounds \cite{porrati02}.\\
The debate on massive gravity is still open and we choose the Fierz-Pauli mass term as the linearised version of the CD theory.\\
Coming back to eq.(\ref{trace}), the choice $\lambda=-2\mu$ makes the trace of the perturbation to vanish in vacuum and now eq.(\ref{divergence}) translates in what would have been the Lorentz gauge in General Relativity. This really simplifies the linearised Einstein tensor that reduces to $-\Box h_{\mu\nu}/2$ so that the equations one finally gets are:
\begin{eqnarray}
\left(\Box +\mu \right) h_{\mu\nu}&=&0\label{1}\\
h^{\mu\nu}_{;\nu}&=&0\label{2}\\
h&=&0.\label{3}
\end{eqnarray}

\subsection{Gravitational waves}

The set of eqs.(\ref{1}, \ref{2}, \ref{3}) represent propagating massive waves and plane waves are solutions of this equations: 
\begin{equation}
h_{\mu \nu }=\alpha _{\mu \nu }e^{i\kappa _{\beta }x^{\beta }}.\label{leonde}
\end{equation} 
Designating the propagation direction as the $z$ axis, the wave vector is $\kappa
=\left( \omega ,0,0,ck\right) $ and from the eq.(\ref{leonde}) we obtain the
dispersion relation:
\[
\omega =\pm c\sqrt{k^{2}-\mu } 
\]

The waves are subluminal, which is commonly referred to as being "massive" ($\mu<0$, consistently with the cosmological limit).
Let us now look at which are the polarisation modes of this spacetime. From eqs.(\ref{1},\ref{2},\ref{3}), we have $5$ dynamical degrees of freedom but this does not mean that the expect the same number of polarisation modes. In a metric theory of gravity there can be at most six polarisation modes, as shown in \cite{eardley}. The analysis can be performed by looking at the geodesic deviation equation, that states which is the displacement between a pair of free-falling particles when a gravitational wave arrives. The three-acceleration depends on the "electric" components of the Riemann tensor ($R^i_{0k0}$). It can be shown that there are six algebraically independent components of the Riemann tensor by using the Newman-Pensore formalism. First, one choose a complex null basis, the so-called null tetrad $(k, l, m, \bar{m})$, that is related to a cartesian system $\{t,x,y,z\}$ by
\begin{eqnarray*}
k&=&\frac{1}{\sqrt{2}}(1,0,0,1),\quad l=\frac{1}{\sqrt{2}}(1,0,0,-1),\\
m&=&\frac{1}{\sqrt{2}}(0,1,i,0),\quad \bar{m}=\frac{1}{\sqrt{2}}(0,1,-i,0).
\end{eqnarray*}
We remember that we have chosen to orient the axes so that the wave travels in the $+z$ direction, and, $u$ being $u=t-z/c$, the $k$ vector is proportional to $\nabla u$.
Then it is possible to split the Riemann tensor into irreducible parts \cite{newman}: the Weyl tensor ($\Psi_0,\Psi_1,\Psi_2,\Psi_3,\Psi_4$), the traceless Ricci tensor ($\Phi_{00},\Phi_{01},\Phi_{11}$, $\Phi_{12},\Phi_{22},\Phi_{02}$, that are  five complex scalars) and the Ricci scalar ($\Lambda$).\\
When considering plane waves only some of them are different from zero and, among these, six are independent. The ones that are helicity $(s)$ eigenstates under rotations about the $z$ axis are
\begin{eqnarray*}
\Psi_2, \Phi_{22}&\rightarrow&\quad s=0,\\
\Psi_3, \bar{\Psi}_{3}&\rightarrow&\quad s=1,\\
\Psi_4, \bar{\Psi}_{4}&\rightarrow&\quad s=2.\\
\end{eqnarray*}
We recall that these six wave amplitudes are observer dependent but there are some invariant statements that are true for all standard observers\footnote{To determine standard observers each observer sees the wave travelling in the z-dir and measures the same frequency for a monochromatic wave.} if they are true for anyone. These statements constitute the $E(2)$ classification of waves, based on the Petrov type of the Weyl tensor \cite{petrov}.

These are related to the Riemann tensor, in cartesian coordinates, as follows \cite{eardley}:
\begin{eqnarray*}
\Psi_2(u)&=&-\frac{1}{6} R_{z0z0}(u),\\
\Psi_3(u)&=&-\frac{1}{2} R_{x0z0}(u)+\frac{i}{2} R_{y0z0}(u),\\
\Psi_4(u)&=&-R_{x0x0}(u)+R_{y0y0}(u)+2i R_{x0y0}(u),\\
\Psi_{22}(u)&=&-\left(R_{x0x0}(u)+R_{y0y0}(u)\right).
\end{eqnarray*}
What we measure in a detection experiment is the relative acceleration of test masses, that is the six "electric" components of the Riemann tensor.  One can express these informations in the so-called {\it driving-force matrix}
$$
S_{ij}(t)=R_{0i0j}.
$$
In general there are eight unknowns, six polarisations and two direction cosines, but if one can establish the direction of the gravitational wave by other information, i.e. the $k$ direction is known, the six elements of the Riemann tensor are sufficient to determine the amplitudes of the gravitational waves. \\
Let us now apply this approach to our theory, where eqs.(\ref{1},\ref{2},\ref{3}) must be satisfied. The second set, applied to a perturbation that propagates in the positive $z$ direction, shows that the $h_{0\mu}$ modes are proportional to the $h_{\mu z}$ ones. The null trace condition, instead, allows us to express the $h_{00}$ or the $h_{zz}=\omega^2 h_{00}/k^2$ to the $h_{xx}$ and $h_{yy}$ modes. In our case we still have all the six polarisation modes, as can be seen by computing the linearised Riemann tensor.\\
Coming back to the driving force matrix, it can be expresse in terms of the basis polarization matrices in the $z$ direction:
$$
S_{ij}(t)=\sum_{r=1}^{6}  p(z,t)^r
 e(z)_{ij}^r,
 $$
 where the amplitudes $p_r(z,t)$ are real and the index $r$ runs over the six modes. 
 The polarization tensor has the form \cite{eardley}:
\begin{eqnarray}\label{polariztensor}
e(z)_{ij}^{1}&=&-6\left(
\begin{array}{ccc}
 0 & 0 & 0 \\
 0 & 0 & 0 \\
 0 & 0 & 1 \\
\end{array}
\right), ~~ e(z)_{ij}^{2}=-2\left(
\begin{array}{ccc}
 0 & 0 & 1 \\
 0 & 0 & 0 \\
 1 & 0 & 0 \\
\end{array}
\right), \nonumber \\
e(z)_{ij}^{3}&=&2\left(
\begin{array}{ccc}
 0 & 0 & 0 \\
 0 & 0 & 1 \\
 0 & 1 & 0 \\
\end{array}
\right),~~e(z)_{ij}^{4}=-\frac{1}{2}\left(
\begin{array}{crc}
1 & 0 & 0 \\
 0 & -1 & 0 \\
 0 & 0 & 0 \\
\end{array}
\right), \nonumber \\
e(z)_{ij}^{5}&=&\frac{1}{2}\left(
\begin{array}{ccc}
 0 & 1 & 0 \\
 1 & 0 & 0 \\
 0 & 0 & 0 \\
\end{array}
\right), ~~ e(z)_{ij}^{6}=-\frac{1}{2}\left(
\begin{array}{ccc}
1 & 0 & 0 \\
 0 & 1 & 0 \\
0 & 0 & 0 \\
\end{array}
\right);
\end{eqnarray}
 the first tensor being related to $\Psi_2$, the second and the third to the real and imaginary part of $\Psi_3$, the two next are those that correspond to $\Psi_4$ and the last one is relative to the scalar $\Phi_{22}$ mode.\\
These modes can play a role  in discriminating among theories of gravity, and in particular they can leave a signature on the CMB anisotropies \cite{bessada,bessada2}

\subsection{Static  weak field limit}
 When we want to reduce to Newtonian limit fields and eventual sources are taken to be static .

Looking at eq.(\ref{1}) we get a screened Poisson equation:
\begin{equation}\label{Yukawa}
\nabla^2 h_{\mu\nu}=-\mu h_{\mu\nu},
\end{equation}
whose solution is a Yukawa-like potential.

\section{Observational constraints}

The only parameter left in our theory is $\mu $. In order to quantify it we
may have recourse to the fit of the type Ia supernovae luminosity which we
presented in \cite{CD}. There we found a value for the bulk modulus;
it was $B\sim 10^{-52}$ m$^{-2}$. From this result we have 
\[
|\mu| \sim 10^{-51}\text{ m}^{-2}.
\]
In order to compare it with the upper bounds that we find in literature it is worthwhile to rewrite our "mass" parameter by using the Planck constant $\hbar$ and the speed of light $c$ so to get it with dimensions of a mass:
$$
m_g=\sqrt{|\mu|}\frac{\hbar}{c}\simeq 6\cdot 10^{-66} \text{kg}.
$$
We know that General Relativity passes all Solar System tests so that this immediately provides an upper limit for the $\mu$ parameter that determines the Yukawa-like fall off of eq.(\ref{Yukawa}) \cite{talmadge}. Moreover, the absence of this effect at the galaxy cluster level provides the limit we were able to find \cite{goldhaber}:
$$
m_g\leq 2\cdot 10^{-65} \text{kg}.
$$
Other limits come from the dispersion in gravitational waves since, if the graviton had a rest mass, the decay rate of an orbiting binary would be affected. As the decay rates of binary pulsars agree very well with GR, the errors in their agreements provide a limit on the graviton mass \cite{damour91,taylor92,finn} but this limit is dramatically less restrictive than the one from the Yukawa potential.
There are a lot of work on similar effects on the timing of a pulsar signal propagating in a gravitational field \cite{baskarav}, or on the measurements of dispersion in gravitational waves using interferometers or by observing gravitational radiation from in-spiralling orbiting binaries \cite{jones,will,larson,cutler,will09}.\\
Finally, an exhaustive review on "massive" gravitons has been done by Goldhaber and Nieto \cite{goldhaber08}.




\end{document}